\documentclass[entropy,pdflatex,article]{mdpi} 

\firstpage{1} 
\pubvolume{1}
\issuenum{1}
\articlenumber{0}
\pubyear{2024}
\copyrightyear{2024}
\datereceived{ } 
\daterevised{ } 
\dateaccepted{ } 
\datepublished{ } 

\hreflink{https://doi.org/} 
\newcommand{\nn}{\nonumber}
\newcommand{\be}{\begin{equation}}
\newcommand{\ee}{\end{equation}}
\newcommand{\bea}{\begin{eqnarray}}
\newcommand{\eea}{\end{eqnarray}}

\Title{Hunting for bileptons at hadron colliders}

\TitleCitation{Hunting for bileptons at hadron colliders}

\Author{Gennaro Corcella $^{1,}$
  \orcidA{}}

\AuthorNames{Gennaro Corcella}

\AuthorCitation{Corcella, G.}

\address{%
$^{1}$ \quad INFN, Laboratori Nazionali di Frascati, Via E.~Fermi 54, 00044, Frascati (RM), Italy; gennaro.corcella@lnf.infn.it}

\corres{Correspondence: gennaro.corcella@lnf.infn.it}

\abstract{  I review possible signals at hadron colliders of bileptons,
  namely doubly-charged vectors or scalars with lepton number $L=\pm 2$,
  as predicted by a 331 model, based on a 
  $SU(3)_C\times SU(3)_L\times U(1)_X$ symmetry.
  In particular, I account for a version of the 331 model, wherein 
  the embedding of the hypercharge is obtained
  with the addition of 3 exotic quarks and vector bileptons. Furthermore,
  a sextet of $SU(3)_L$, necessary to give masses to leptons, yields an extra
  scalar sector, including a doubly-charged Higgs, i.e. scalar bileptons.
  As bileptons are mostly produced in pairs at hadron colliders, their main
  signal is given by two same-sign lepton pairs at high invariant mass.
  Nevertheless, they can also decay according to  non-leptonic modes,
  such as a TeV-scale heavy quark, charged 4/3 or 5/3, plus a Standard Model
  quark. I explore both leptonic and non-leptonic decays and the
sensitivity to such processes of present and future hadron colliders.}


\begin{document}

\section{Introduction}
The Standard Model (SM) of electroweak and strong interactions is a complete
theory, but it exhibits several drawbacks, such as the hierarchy problem in the Higgs sector, neutrino masses or Dark Matter, which call for a
theory with a more general gauge structure and possibly new particles.
As well motivated SM extensions, such as supersymmetry or extra dimensions,
have so far given no visible signal at the LHC, it is mandatory exploring
alternative scenarios. In this paper, I review work carried
out in the latest few years \cite{cccf1,cccf2,cccf3} in the framework of the
$SU(3)_C\times SU(3)_L\times U(1)_X$ model \cite{PHF,PP,Valle},
also known as 331 model, and its possible signals at the Large
Hadron Collider
(LHC) and at a future 100 TeV hadron collider (FCC-$hh$).

Among its main features, this model predicts the existence of bileptons,
i.e. gauge bosons $(Y^{--}, Y^{++})$ of charge $Q=\pm 2$ and lepton number
$L=\pm 2$, which is why one often refers to it as bilepton model.
Furthermore, in the specific formulation of \cite{PHF}, 
one is capable of explaining the asymmetry of the third
quark family, i.e. top and bottom quarks, with respect to the other two,
while the existence of three families, i.e. $N_f=N_C=3$,
$N_C$ being the number of colours, is a consequence of the requirement of an
anomaly-free theory (see also the detailed discussion in \cite{melle}). 

As will be detailed later on, the scenario which will be investigated,
besides the vectors $Y^{++(--)}$, predicts a number of new particles, which may
possibly be in the reach of the LHC or a future hadron collider, such as
FCC-$hh$. Among those, one has heavy quarks with charge 5/3, usually labelled
$T$, or charge 4/3, i.e. $D$ or $S$, which typically have a mass of the order
of few TeV (see the analysis in \cite{cccf3}). Moreover, a complete
description of the model requires the inclusion of a Higgs sector, which
is a sextet of $SU(3)_L$ and is needed to give mass to the leptons.
In the Higgs sector, the predicton of
new doubly-charged scalars is particularly relevant.
Such Higgs-like bosons with charge
$\pm 2$ have been intensively searched by the experimental collaborations
in different new physics models,
setting mass bounds between 900 and 1100 GeV \cite{atlhpp,cmshpp} at
$\sqrt{s}=13$~TeV and integrated luminosities ${\cal L}=139$~fb$^{-1}$ and 12.9~fb$^{-1}$,
respectively. As far as I know, no specific search for vector
bileptons has been undertaken so far.

On the other hand, as will be detailed in the following,
Refs.~\cite{cccf1,cccf3} investigated the phenomenology of vector bileptons,
decaying into leptonic or non-leptonic final states, while Ref.~\cite{cccf2}
explored both vector and scalar bileptons, concentrating on final states
with same-sign lepton pairs. All such papers published results
for reference points which are not yet excluded by the experimental
searches, with a bilepton mass just below the exclusion range, in order
to maximize the production cross section.

The plan of this contribution is the following.
In Section 2 I shall review the main ingredients of the 331 model, in the
version proposed in \cite{PHF}. In Section 3 I shall critically
present the phenomenological results contained in Refs.~\cite{cccf1,cccf2,cccf3}. In Section 4 some concluding remarks will be made.

\section{Theoretical framework}

Following Ref.~~\cite{PHF}, the gauge structure of the bilepton model is 
$SU(3)_c \times SU(3)_L \times U(1)_X$, with the fermions (quarks) in the fundamental of $SU(3)_c$ arranged into
triplets of $SU(3)_L$. As anticipated in the introduction,
the third quark family (top and bottom) is treated 
asymmetrically with respect to the first two families in the electroweak 
$SU(3)_L$. In detail, as for the first two families, one has  
\begin{equation}
Q_1=\left(
\begin{array}{c}
u_L\\
d_L\\
D_L
\end{array}
\right),\quad Q_2=\left(
\begin{array}{c}
c_L\\
s_L\\
S_L
\end{array}
\right),\quad Q_{1,2}\in(3, 3, -1/3)
\end{equation}
under $SU(3)_c \times SU(3)_L \times 
U(1)_X$,
while, for the third one, it is
\begin{equation}
Q_3=\left(
\begin{array}{c}
b_L\\
t_L\\
T_L
\end{array}
\right),\quad Q_3\in(3,\bar3, 2/3).
\end{equation}
In the above formulation, $D$, $S$ and $T$ are quarks,
with charge 4/3 ($D$ and $S$) or 5/3 ($T$). In the following, 
I will explore scenarios wherein such quarks are either within or
outside the reach of present and future hadron colliders.

The right-handed quarks ($\bar q$), as happens in 
the SM, are singlets even under $SU(3)_L$. Their representations are the
following: 
 \begin{align}
( d_R,  s_R, b_R)&\in (\bar 3, 1, 1/3)\\
( u_R,  c_R,  t_R) &\in(\bar 3, 1, -2/3)\\
( D_R,  S_R) &\in (\bar 3, 1, 4/3)\\
 T_R &\in (\bar 3, 1, -5/3).
 \end{align}
 One can notice that adding such new particles to the Standard Model states
 is not enough to cancel the $SU(3)_L$ anomalies \cite{PHF,melle}.
 Therefore, one has to
 introduce new leptonic states in three $\bar{3}$ representation.
 As a result, the three lepton families, unlike the quarks, are arranged in
 a `democratic' manner as triplets of $SU(3)_L$:
\begin{equation}
l=\left(
\begin{array}{c}
l_L\\
\nu_l\\
\bar l_R
\end{array}
\right),\quad l\in(1, \bar 3, 0),\quad l=e,\ \mu,\ \tau.
\end{equation}
As discussed in \cite{PHF,melle} these assignments of quarks and leptons
lead to the cancellation of the anomaly of $SU(3)_L$, while the
$SU(3)_C$ one is cancelled as happens in the SM, i.e. through a complete 
balance between left-handed colour triplets and right-handed anti-triplets
in the quark sector.

The electroweak symmetry breaking of this 331 model occurs through
scalar fields $\rho$, $\eta$ and $\chi$
which are arranged as triplets of $SU(3)_L$:
\begin{equation}
\rho=\left(
\begin{array}{c}
\rho^{++}\\
\rho^+\\
\rho^0
\end{array}
\right)\in(1,3,1),\quad\eta=\left(
\begin{array}{c}
\eta^+\\
\eta^0\\
\eta^-
\end{array}
\right)\in(1,3,0),\quad\chi=\left(
\begin{array}{c}
\chi^0\\
\chi^-\\
\chi^{--}
\end{array}
\right)\in(1,3,-1).
\end{equation}
The breaking of $SU(3)_L\times U(1)_X\to U(1)_{\rm{em}}$ is achieved in two
steps. First, 
the vacuum expectation value (vev) of the neutral component of $\rho$
gives mass to the novel gauge bosons, $Z'$, $Y^{++}$ and $Y^+$ and
heavy quarks 
$D$, $S$ and $T$. In this first step, 
the original gauge group $SU(3)_L\times U(1)_X$ breaks into
$SU(2)_L\times U(1)_Y$. In the second step, it is $\chi^0$ and $\eta^0$ 
which get a vev and one has 
the usual breaking from $SU(2)_L\times U(1)_Y$ to $U(1)_{\rm{em}}$.

In detail, the scalar potential reads: 
\begin{align}
V&= m_1\, \rho^*\rho+m_2\,\eta^*\eta+m_3\,\chi^*\chi\nn\\
&\quad+\lambda_1 (\rho^*\rho)^2+\lambda_2(\eta^*\eta)^2+\lambda_3(\chi^*\chi)^2\nn\\
&\quad +\lambda_{12}\rho^*\rho\,\eta^*\eta+\lambda_{13}\rho^*\rho\,\chi^*\chi+\lambda_{23}\eta^*\eta\,\chi^*\chi\\
&\quad +\zeta_{12}\rho^*\eta\,\eta^*\rho+\zeta_{13}\rho^*\chi\,\chi^*\rho+\zeta_{23}\eta^*\chi\,\chi^*\eta\nn\\
&\quad +\sqrt2 f_{\rho\eta\chi} \rho\, \eta\, \chi\nn . 
\end{align}
The neutral component of each triplet acquires a vev and can be expanded as
\bea
\rho^0&=&\frac{1}{\sqrt2}v_\rho+\frac{1}{\sqrt2}\left({\rm Re}\,\rho^0+i{\rm Im}\rho^0\right)\\
\eta^0&=&\frac{1}{\sqrt2}v_\eta+\frac{1}{\sqrt2}\left({\rm Re}\,\eta^0+i{\rm Im}\eta^0\right)\\
\chi^0&=&\frac{1}{\sqrt2}v_\chi+\frac{1}{\sqrt2}\left({\rm Re}\,\chi^0+i{\rm Im}\,\chi^0\right).
\eea
As detailed in \cite{cccf1}, one first determines 
the potential minimization conditions and then, after spontaneous symmetry breaking, the gauge and the mass eigenstates of $\rho$, $\eta$ and $\chi$.
The explicit expression of the mass matrices of the scalar sector, both neutral and charged, are given in \cite{cccf1} and we do not report them here for the
sake of brevity.

As anticipated in the introduction and discussed in detail in
\cite{cccf2}, it is necessary to add to the scalar sector a
$SU(3)_L$ sextet, in order to give masses to leptons. 
This implies that the particle spectrum of the bilepton model includes
doubly-charged Higgs bosons ($H^{\pm\pm}$) capable of decaying into
same-sign lepton pairs. In other words, decays like
$H^{\pm\pm}\to l^\pm l^\pm$ would be an evidence of the presence of
sextet representation of $SU(3)_L$.

Still on decays of doubly-charged scalars, as pointed out in \cite{cccf2},
in principle, as for the Standard Model Higgs, one should have amplitudes
proportional to the Yukawa coupling, hence to the masses of the final-state
particles. However, for the sake of generality and putting vector and
scalar bileptons on the same footing, following \cite{cccf2} I shall
consider a scenario where the branching ratios of doubly-charged Higgs bosons
are not proportional to the mass, but, referring, e.g., to decays
into same-sign lepton pairs, one has ${\rm BR}(Y^{\pm\pm}\to l^{\pm}l^\pm)\simeq {\rm BR}(H^{\pm\pm}\to l^{\pm}l^\pm)$, $Y^{\pm\pm}$ being vector bileptons.

After electroweak symmetry breaking, one ends up with a rich Higgs sector.
In detail, we have 5 scalar Higgs bosons, one of them being
the Standard Model one with mass about 125 GeV, 4 neutral pseudoscalar Higgs bosons, out of which 2 are the Goldstones of the $Z$ and the $Z^\prime$ massive vector bosons. Furthermore, one has 6 charged Higgses, 2 of which are the
charged Goldstones and 3 are doubly-charged Higgses, 1 of which is a Goldstone boson.

As the main goal of this investigation is the phenomenology of doubly-charged
vectors and scalars, we point out that
Ref.~\cite{cccf2}
contains a thorough discussion of the vertices where pairs
$Y^{\pm\pm}Y^{\pm\pm}$ or $H^{\pm\pm}H^{\pm\pm}$ are involved. We do not report the formulas in the
present contribution for brevity and refer to \cite{cccf2} for such couplings.

\section{Phenomenology at LHC and future colliders}

\subsection{Leptonic decays of bileptons}

\begin{figure}[H]\begin{center}
  \includegraphics[width=0.2\textwidth]{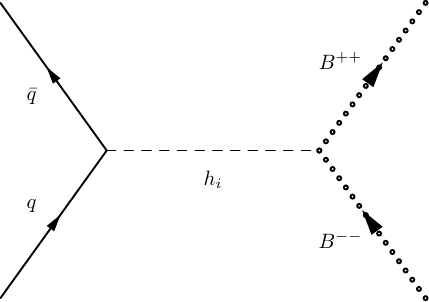}\hspace{.5cm}
  \includegraphics[width=0.2\textwidth]{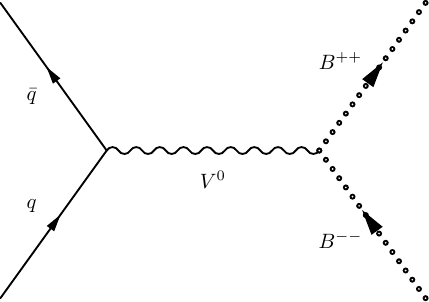}\hspace{.5cm}
  \includegraphics[width=0.2\textwidth]{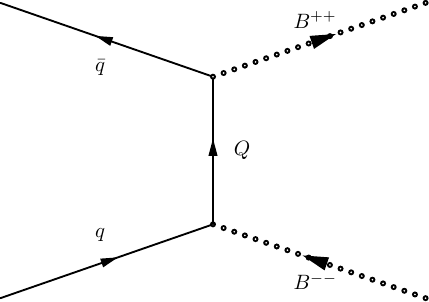}                  \caption{Characteristic diagrams for the production
    of bilepton pairs in hadron collisions.}
    \label{fig1}
\end{center}
  \end{figure}

In this section, I present the main results contained in
\cite{cccf1,cccf2,cccf3} regarding the phenomenology of doubly-charged
scalars and vectors at LHC (13 or 14 TeV) and future colliders, namely FCC-$hh$.
Typical contribution to bilepton production in hadron collisions are
presented in Fig.~\ref{fig1}: an initial-state $q\bar q$ pair annihilates
and a $B^{++}B^{--}$ pair, $B$ being a doubly-charged vector or scalar,
is prduced.
As can be seen, bilepton-pair production can be mediated by the
exchange of, e.g., a neutral Higgs or a vector ($Z$, $Z'$ or photon) in the
$s$-channel, or a heavy TeV-scale quark, charged 5/3, in the $t$-channel.
Ref.~\cite{cccf1} also discusses the production of bilepton pairs in
association with jets and presents some typical diagrams for such
processes as well.
\unskip
In detail, Ref.~\cite{cccf1} and \cite{cccf2} account for decays of bileptons
in same-sign lepton pairs, say $Y^{++}\to \mu^+\mu^+$,
while \cite{cccf3} deals with non-leptonic decays,
i.e. decays into a light (SM) quark (antiquark) and a heavy Tev-scale antiquark
(quark), e.g., $Y^{++}\to T\bar b, \bar Du$.
The results are presented
for a few benchmarks, determined in such a way to be not yet excluded
by the experimental searches, though capable of yielding a remarkable
cross section and number of events. 

In order to determine the benchmarks and scan the parameter space, we
had to implement the bilepton model in the SARAH 4.9.3 code \cite{sarah}.
In particular, Refs.~\cite{cccf1,cccf2} carry out a phenomenological
investigation for a bilepton mass about 880 GeV, just below the
experimental exclusion limit.
In Ref.~\cite{cccf2}, where the phenomenology of vector and
scalar bileptons is compared, one sets the masses to the same value:
\begin{equation}
  M_{Y^{++}}\simeq M_{H^{++}}\simeq 878.3~{\rm GeV},
  \end{equation}
while exotic Higgs bosons, $Z'$ and heavy quarks are assumed to have masses
well above 1 TeV, hence they are too heavy to contribute to any
bilepton phenomenology (see \cite{cccf2} for their actual values in
the reference points).
One can then explore the process
\begin{equation}\label{hhyy}
  pp\to Y^{++}Y^{--}(H^{++}H^{--})\to (l^+l^+)(l^-l^-),
  \end{equation}
setting the following cuts on final-state lepton
transverse momentum ($p_T$), rapidity
($\eta$) and invariant opening angle: 
\begin{equation}
  p_{T,l}>20~{\rm GeV},   |\eta_l|<2.5, \Delta R_{ll}>0.1.
\end{equation}
In \cite{cccf1,cccf2} one assumed democratic leptonic branching ratios
of bileptons, namely\\ ${\rm BR}(Y^{++}\to l^+l^+)\simeq
{\rm BR}(H^{++}\to l^+l^+)\simeq 1/3$ for any lepton flavour
($e$, $\mu$ or $\tau$).

After the cuts are applied, the leading-order cross sections of
processes in Eq.~(\ref{hhyy}), computed by means of MadGraph 2.6.1
\cite{madgraph}
at $\sqrt{s}=13$~TeV,
read:
\be\sigma(pp\to YY\to 4l)\simeq
4.3~{\rm fb}\ ;\ \sigma(pp\to HH\to 4l)\simeq 0.3~{\rm fb} .\ee
At 14 TeV, one has instead: $\sigma(pp\to YY\to 4l)\simeq
6.0$~fb and $\sigma(pp\to HH\to 4l)\simeq 0.4$~fb. 
As discussed in \cite{cccf2},
the higher cross section in the case of vector-pair
production can be explained in terms of the bilepton helicity. 
In the case of doubly-charged Higgs production, only the amplitudes where
the intermediate vectors ($\gamma$, $Z$, $Z'$) have
helicity zero contribute, while, in case of doubly-charged
vectors, all helicities 0 and $\pm 1$ play a role.
For processes mediated by scalars, 
$Y^{++}$ and $Y^{--}$ can still rearrange their helicities in a few different
ways to achieve angular-momentum conservation and a total vanishing
helicity in the centre-of-mass frame. Similar results were also
found in \cite{ramirez}, where the authors investigate vector- and
scalar-bilepton pairs at hadron colliders, at parton level in the LO
approximation.

\begin{figure}[H]
  \begin{center}
  \includegraphics[width=0.4\textwidth]{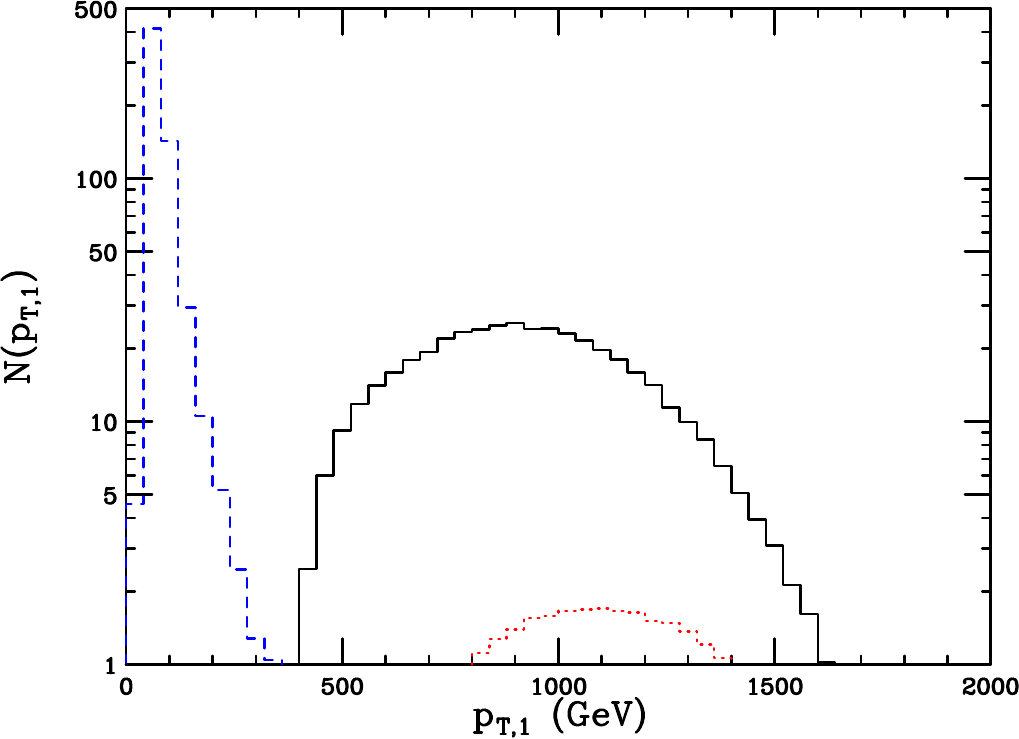}\hspace{1.cm}
  \includegraphics[width=0.4\textwidth]{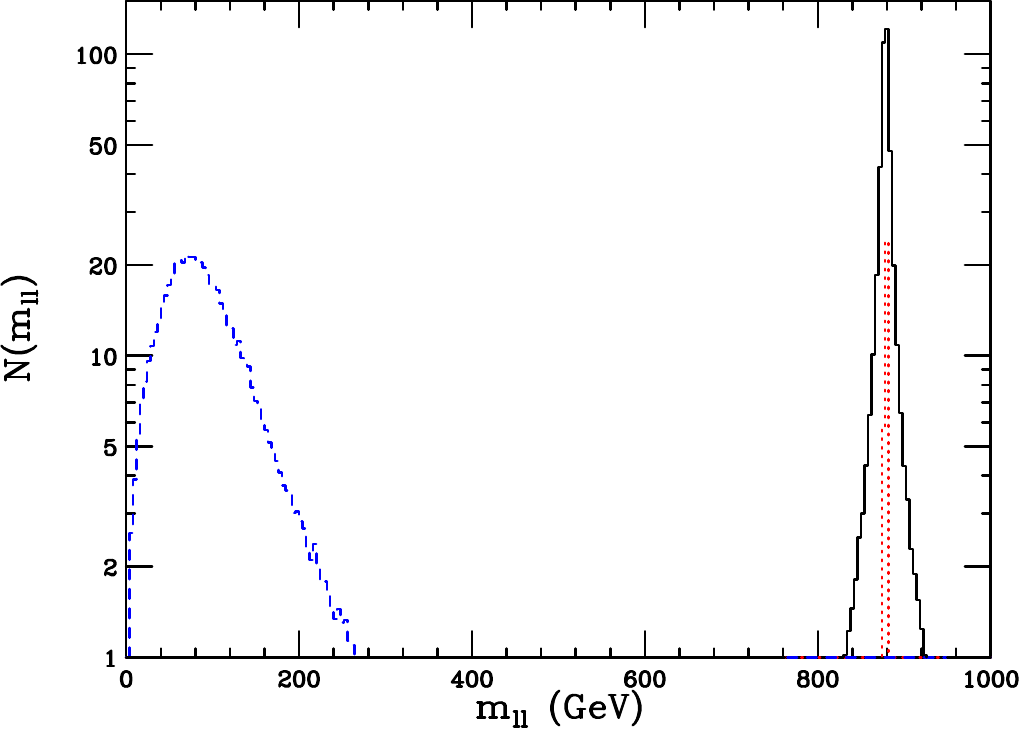}
  \caption{Distributions of the transverse momentum of the hardest lepton
    (left) and of the same-sign lepton invariant mass (right).
    The solid histograms are the spectra yielded by vector bileptons,
  the dots correspond to scalar doubly-charged Higgs bosons,
the blue dashes to the $ZZ$ Standard Model background.}
  \label{fig2}
  \end{center}
  \end{figure}

As for backgrounds, as pointed out in \cite{cccf1,cccf2}, the main
one is due to same-sign lepton-pair production mediated by a $Z$-boson pair,
i.e. \be
pp\to ZZ\to (l^+l^-)(l^+l^-),
\label{zz}
\ee
while processes mediated by neutral-Higgs pairs are negligible due to
the tiny coupling of the Higgs with leptons. 
After setting the cuts, 
the LO cross section of the process (\ref{zz}) is given by
$\sigma(pp\to ZZ\to 4l)\simeq 6.1$~fb at 13 TeV and 6.6 fb at 14 TeV.
For an integrated luminosity ${\cal L}=300$~fb$^{-1}$, at 13 TeV one has $N(YY)\simeq 1302$
lepton pairs mediated by doubly-charged vectors, while scalars yield
$N(HH)\simeq 120$ and the $ZZ$ background $N(ZZ)\simeq 1836$ events.
At 14 TeV and  ${\cal L}=3000$~fb$^{-1}$, such numbers read
$N(YY)\simeq 17880$, $N(HH)\simeq 1260$ and $N(ZZ)\simeq 19740$.
For $S$ signal and $B$ background events,
one can define a significance (in units of standard deviations)  
\be\label{sign}
s=\frac{S}{\sqrt{B+\sigma_B^2}},\ee
where $\sigma_B$ is the systematic error on $B$ gauged
about $\sigma_B\simeq 0.1 B$
in \cite{cccf2}. Following \cite{lista}, the denominator of the significance
(\ref{sign}) sums in quadrature the intrinsic statistical fluctuation of the
background $\sqrt{B}$ and the uncertainty in the background $\sigma_B$,
obtaining $s=S/\sqrt{\sqrt{B^2}+\sigma_B^2}$.
One can then find a significance $s\simeq 6.9$ for vector pairs at
13 TeV and ${\cal L}=300$~fb$^{-1}$ and $s=0.6$ for scalars, which clearly
means that only doubly-charged vector bileptons may possibly be
visible at 13 TeV. At 14 TeV and high integrated luminosity, one has
$s\simeq 9$ for $Y^{++}Y^{--}$ and $s\simeq 0.64$ for $H^{++}H^{--}$
production.
Reference~\cite{cccf2} explores several distributions of relevant
leptonic observables, yielded by vector and scalar bileptons, as well
as $ZZ$ background. For the sake of conciseness, we
present in Fig.~\ref{fig2} only those referring to
the hardest-lepton transverse momentum $p_{T,1}$ and
the same-sign lepton invariant mass.
  As for the transverse momentum $p_{T,1}$,
the $ZZ$ distribution is sharp and peaked at low $p_T$, while
those yielded by the $HH$ and $YY$ bileptons are much broader and
peak at about 1 TeV.
In fact, the background
$Z$ bosons are much lighter than bileptons and
decay into different-sign lepton pairs,
while $Y^{\pm\pm}$ and $H^{\pm\pm}$ decay
into same-sign electrons and muons.
Furthermore, for every value of
$p_T$, the $HH$ spectrum is well below the $YY$ one.

Regarding the same-sign lepton invariant mass $m_{ll}$,
as expected, the 331 signal peaks at
$m_{ll}\simeq 900$~GeV, while the $Z$-background distribution is
instead a broad spectrum, significant up to about 350 GeV
and maximum around 70 GeV. The signal spectra are pretty narrow:
the authors of \cite{cccf2} quoted $Y^{++}$ and $H^{++}$ widths
about 7 GeV and 400 MeV, hence much smaller that their masses.

Before concluding this subsection, one can then point out that, as should have
been expected due to the obtained significances, the distributions
in Fig.~\ref{fig2}, as well as those published in \cite{cccf2},
seem to show that discriminating the 331 signal from the background
should be feasible, with doubly-charged vectors dominating over scalars.
  
\subsection{Non-leptonic decays of bileptons}
While possible decays into same-sign lepton pairs would be the
`smoking gun' for bilepton discovery at the LHC, depending on the mass
spectrum, it is also possible that vector and
scalar bileptons could well decay into non-leptonic final states, such as a
TeV-scale
heavy quark and a light quark. This was in fact the main purpose of the
exploration in \cite{cccf3}, which I shall summarize hereafter.

Unlike Refs.~\cite{cccf1,cccf2}, the more recent work in \cite{cccf3}
took advantage of the results of Ref.~\cite{paulcla}, where the authors,
by using
renormalization group arguments, gave the estimate $m_Y=(1.29\pm 0.06)$~TeV
for the bilepton mass.
Making use of this finding, Ref.~\cite{cccf3} concentrated on
doubly-charged vectors $Y^{\pm\pm}$ and
considered two benchmark cases: one scenario with all heavy quarks
$D$, $S$ and $T$ lighter than $Y^{\pm\pm}$ and another one  
where only the mass of $D$ is lower than $m_Y$, while $S$ and $T$ are
heavier.
More precisely, the first benchmark, labelled BM I in \cite{cccf3}
sets all TeV-scale quark masses to 1 TeV, i.e.
\be m_D=m_S=m_T=1~{\rm TeV},\ee
while in the second one, i.e. BM II, one has the following mass values:
\be m_D=1.2~{\rm TeV}, \ m_S=1.5~{\rm TeV}, \ 
m_T=1.5~{\rm TeV}.\ee
Both BM I and BM II are consistent with a light SM-like Higgs boson with
mass about 125 GeV; all the other BSM particles have mass much above 1 TeV,
therefore they are not relevant for bilepton phenomenology.

Unlike Refs.~\cite{cccf1,cccf2}, wherein bileptons could only decay leptonically
in the benchmark points of \cite{cccf3} one has substantial branching
fractions into both leptonic and hadronic final states.
In detail, one has
\be   {\rm BR}(Y^{++}\to l^+ l^+)\simeq 20.6\%\ ({\rm BM\ I}),\ 32.5\% ({\rm BM\ II}),
\end{equation}
for each lepton flavour
$l=e,\mu,\tau$, and
\begin{equation}\label{decy2}
  {\rm BR}(Y^{++}\to u\bar D, c\bar S, T\bar b)\simeq 12.7\%\ ({\rm BM\ I}),\ 
  {\rm BR}(Y^{++}\to u\bar D)\simeq 2.5\%\  ({\rm BM\ II}).
  \end{equation}
The total bilepton widths instead read:
\begin{equation}\label{wy}
  \Gamma(Y^{\pm\pm})\simeq 17.9~{\rm GeV}\ ({\rm BM\ I});\
  \Gamma(Y^{\pm\pm})\simeq 11.4~{\rm GeV}\ ({\rm BM\ II}).
  \end{equation}
The larger width in BM I is clearly due to the fact that decays into
final states with all three heavy quarks $D$, $S$ and $T$ are permitted.

Before presenting some numerical results, one should also explore the
phenomenology of TeV-scale quark decays.
In BM I, the heavy
quarks exhibit three-body decays into a Standard Model quark
and a same-sign lepton pair or a lepton-neutrino pair, 
through a virtual bilepton, with the following branching fractions:
\begin{equation}\label{decay1}
  {\rm BR} (D (S)\to u (c)  l^-l^-)\simeq
  {\rm BR} (D (S)\to d (s) l^-\nu_l) \simeq 16.7\%\ 
({\rm BM\ I}).
\end{equation}
In BM II, $S$ and $T$ are heavier
than singly- and doubly-charged bileptons and can therefore decay
into final states with a real $Y^\pm$ or $Y^{\pm\pm}$.
While the $D$ rates are the same as in BM I, i.e. Eq.~(\ref{decay1}), 
$S$ can decay into real bileptons as follows:
\begin{equation}\label{decay2}
  {\rm BR} (S\to c Y^{--})\simeq 50.5\%,\ 
  {\rm BR} (S\to  sY^-)\simeq 49.5\%\ 
({\rm BM\ II}).
\end{equation}
As for $T$, charged 5/3, its decay rates are:
\bea\label{dect1}
  {\rm BR} (T\to b l^+l^+)\simeq 19.4\%,\ 
  {\rm BR} (T\to t l^+\bar \nu_l)\simeq 13.9\%\  ({\rm BM\ I});\\
  {\rm BR} (T\to b Y^{++})\simeq 64.6\%,\  
  {\rm BR} (T\to t Y^+)\simeq 35.4\%\  ({\rm BM\ II}).
  \eea
The total decay widths are given by:
\bea\label{wq1}
  \Gamma(D)\simeq \Gamma (S) \simeq 3.4\times 10^{-3}~{\rm GeV},\
  \Gamma(T)\simeq 3.0\times 10^{-3}~{\rm GeV}\ ({\rm BM\ I});\\
  \Gamma(D)\simeq 1.3\times 10^{-2}~{\rm GeV},\ 
  \Gamma (S) \simeq 1.5~{\rm GeV},\
  \Gamma(T)\simeq 1.1~{\rm GeV}\ ({\rm BM\ II}).
  \eea
In other words, in BM I all TeV-scale quarks have a pretty small width, of
the order  ${\cal O}(10^{-3}~{\rm GeV})$; in BM II 
$D$ is still quite narrow, having a width ${\cal O}(10^{-2}~{\rm GeV})$, 
while the widths of $S$ and $T$ are of the order of 1 GeV, since they
are capable of decaying
into states with real bileptons.

The production cross section of bilepton pairs at LHC (13 and 14 TeV)
and FCC-$hh$ are given by:
\bea\label{xlhc}
  \sigma(pp\to Y^{++}Y^{--}) \simeq 0.75~{\rm fb}\ ({\rm LHC}, 13~{\rm TeV}),\\
  \sigma(pp\to Y^{++}Y^{--}) \simeq 1.12~{\rm fb}\ ({\rm LHC}, 14~{\rm TeV}),\\
  \sigma(pp\to Y^{++}Y^{--}) \simeq 393.89~{\rm fb}\ ({\rm FCC-}hh),\eea
with the FCC-$hh$ cross sections about 500 and 350
times larger than the LHC ones.

Following \cite{cccf3}, in BM I I shall account for 
primary decays of $Y^{\pm\pm}$  into quarks $T$, which further
decay into a bottom quark and a same sign muon pair,
hence a final state with four $b$-flavoured jets and two same-sign
muon pairs:
\begin{equation}\label{dec1}
  pp \to Y^{++}Y^{--}\to (T\bar b)(\bar T b)\to (b\bar b \mu^+\mu^+)
(b\bar b\mu^-\mu^-)
  \ ({\rm BM\ I}).\end{equation}
In reference point BM II, I shall instead explore
primary decays into quarks $D$ and 
final states with
four $u$-quark initiated light jets accompanied by four muons
($4u4\mu$):
\begin{equation}\label{dec2}
  pp \to Y^{++}Y^{--}\to (\bar D u )(D\bar u)\to (u\bar u \mu^+\mu^+)
(u\bar u\mu^-\mu^-)
  \ ({\rm BM\ II}).\end{equation}
In Ref.~\cite{cccf3} a few representative diagrams of processes
(\ref{dec1}) and (\ref{dec2}) are presented as well.

A first rough estimation of the predicted number of
events at LHC and FCC-$hh$
can be obtained by multiplying the inclusive cross sections
in Eqs.~(\ref{xlhc}) by the relevant branching ratios,
assuming a perfect tagging efficiency and no cut on final-state
jets and leptons. At LHC one obtains:
\bea\label{fulllhc}
  \sigma(pp\to YY\to 4b4\mu)\simeq 4.55\times 10^{-4}~{\rm fb}\ ({\rm LHC,
    13~TeV,\ BM\ I}),\\
  \sigma(pp\to YY\to 4b4\mu)\simeq 6.80\times 10^{-4}~{\rm fb}\ ({\rm LHC,
    14~TeV,\ BM\ I}),\\
  \sigma(pp\to YY\to 4u4\mu)\simeq 1.31\times 10^{-5}~{\rm fb}~({\rm LHC, 13~TeV,\ BM\ II}),\\
  \sigma(pp\to YY\to 4u4\mu)\simeq 2.03\times 10^{-5}~{\rm fb}~({\rm LHC, 14~TeV,\ BM\ II}).
  \eea
Such cross sections are too small to see any event at 300~fb$^{-1}$ and 
at 3000~fb$^{-1}$ (HL-LHC), even before imposing any acceptance cut.
Therefore, the investigation in \cite{cccf3} discarded the LHC environment
and the analysis was concentrated on FCC-$hh$, where the cross sections
are remarkable:
\bea
\sigma(pp\to YY\to 4b4\mu)\simeq 0.24~{\rm fb}\ ({\rm FCC-}hh,\ {\rm BM\ I}),
\\
  \sigma(pp\to YY\to 4u4\mu)\simeq 6.87\times 10^{-3}~{\rm fb}~({\rm FCC}-hh,\
        {\rm BM\ II}).
  \eea
The scenario BM I at FCC-$hh$ yields a few hundreds events; BM II is less
promising, but still worthwhile to investigate.

As for the backgrounds, one considers, above all, four $b$ quarks and two $Z$ bosons decaying into muon pairs (background $b_1$),
\begin{equation}\label{back1}
  pp\to bb\bar b\bar bZZ\to  bb\bar b\bar b\mu^+\mu^-\mu^+\mu^-,
\end{equation}
and four top quarks
with the subsequent $W$'s decaying into
muons and requiring, as in \cite{cccf2}, a small missing energy
due to the muon neutrinos (background $b_2$):
\begin{equation}\label{back2}
  pp\to t t \bar t\bar t\to (bW^+)(bW^+)(\bar bW^-)(\bar bW^-)\to
  bb\bar b \bar b \mu^+\mu^+\mu^-\mu^-\nu_\mu\nu_\mu\bar\nu_\mu\bar\nu_\mu.
\end{equation}
As discussed in \cite{cccf3}, the simulation of (\ref{back2}) accounted
for electroweak corrections as well, since, as pointed out in \cite{zaro},
at both LO and NLO they can contributed up to 10\% of the total cross section.

Reference~\cite{cccf3} also considered the following backgrounds
with four light jets and two $Z$ bosons and with two light jets,
two $b$-jets and two $Z$'s:
\bea\label{back3}
  pp\to jjjjZZ\to jjjj\mu^+\mu^-\mu^+\mu^-,
  pp\to jjb\bar bZZ\to jjb\bar b\mu^+\mu^-\mu^+\mu^-.
\eea
In Eqs.~(\ref{back3})
$j$ is either a light-quark or gluon-initiated jet, mistagged as a $b$-jet.

I cluster the final-state hadrons in four jets
and apply the following acceptance cuts on jets and muons:
\begin{eqnarray}
\label{cuts}
&& p_{T,j}>30~{\rm GeV},\  p_{T,\mu}>20~{\rm GeV},\  |\eta_j|<4.5,
|\eta_\mu|<2.5,\nonumber\\
&& \Delta R_{jj}>0.4,\   \Delta R_{\mu\mu}>0.1,\ 
\Delta R_{j\mu}>0.4,\  {\rm MET}<200~{\rm GeV}.
\end{eqnarray}
The cuts in (\ref{cuts}) correspond to a conservative choice of the so-called
`overlap removal' algorithm used at the LHC to discriminate lepton and jet tracks at LHC \cite{removal}. As for the four-top background (\ref{back2}), Ref.~\cite{cccf3} sets the additional cut ${\rm MET}<200$~GeV on the missing transverse 
energy due to the neutrinos in the final state. In \cite{cccf3}
the MET cut was consistently set even on neutrinos coming from
hadron decays.

In principle, one should account for the $b$-tagging
efficiency, as well as the probability of mistagging a light jet as a $b$-jet.
Such efficiencies depend on the jet rapidity, transverse momentum and flavour;
however, for an explorative analysis,
like the one in \cite{cccf3}, one can 
implement such effects in a flat manner, i.e. independently
of the jet kinematics and of the flavour of the light jets.
The $b$-tagging efficiency ($\epsilon_b$) and
and the mistag rate ($\epsilon_j$) and the mistag rate ($\epsilon_j$, with $j=u,d,s,c$) are then set to the following values, as in \cite{delphes}:
\begin{equation}\label{tag}
  \epsilon_b=0.82\ ,\ \epsilon_j=0.05.
\end{equation}

After setting such cuts, the signal ($s$) cross section of process
(\ref{dec1}) 
  amounts to $\sigma(4b4\mu)_s\simeq 6.24\times 10^{-2}$~fb, leading to
$N(4b 4\mu)_s\simeq 90$ events
  at FCC-$hh$ for an integrated luminosity ${\cal L}=3000$~fb$^{-1}$ and after setting
  all cuts and $b$-tagging efficiency.
As for the backgrounds (\ref{back1})--(\ref{back3}),
one gets
$\sigma(4b4\mu)_{b_1}\simeq 1.28\times 10^{-2}$~fb,
$\sigma(4b4\mu+{\rm MET})_{b_2}\simeq 3.34\times 10^{-2}$~fb,
$\sigma(4j4\mu)_{b_3}\simeq 4.43$~fb,
$\sigma(2b2j4\mu)_{b_4}\simeq 1.34$~fb.

Including also $b$-tagging and mistag
efficiencies and rounding to the nearest ten, one computes the following
number of background events at FCC-$hh$:
$N(4b4\mu)_{b_1}\simeq 20$,
$N(4b4\mu+{\rm MET})_{b_2}\simeq 50$. The
backgrounds $b_3$ and $b_4$ yield too few events to be significant.

Regarding BM II and the decay chain (\ref{dec2}), the cross section
is about $\sigma(4j4\mu)_s\simeq 1.88\times 10^{-3}$~fb at
FCC-$hh$.
As a result, considering that some extra suppression is due
to the efficiency of jet/lepton tagging, the BM II reference point
was eventually discarded in \cite{cccf3}.

\begin{figure}[H]\begin{center}
  \includegraphics[width=0.47\textwidth]{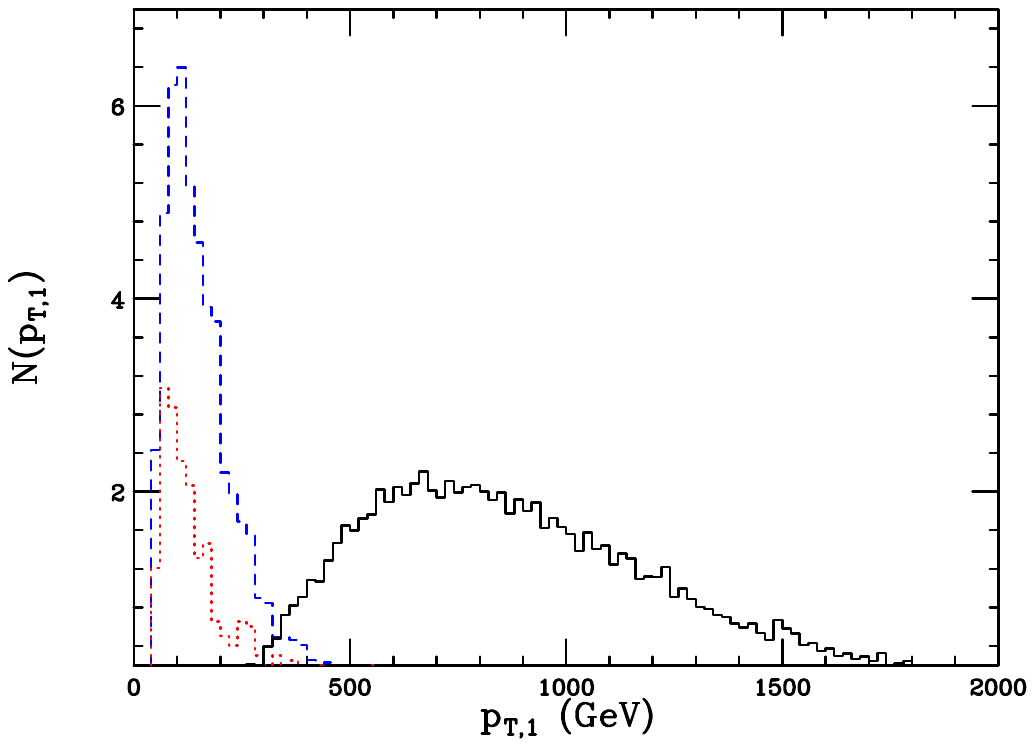}\hspace{0.5cm}
  \includegraphics[width=0.47\textwidth]{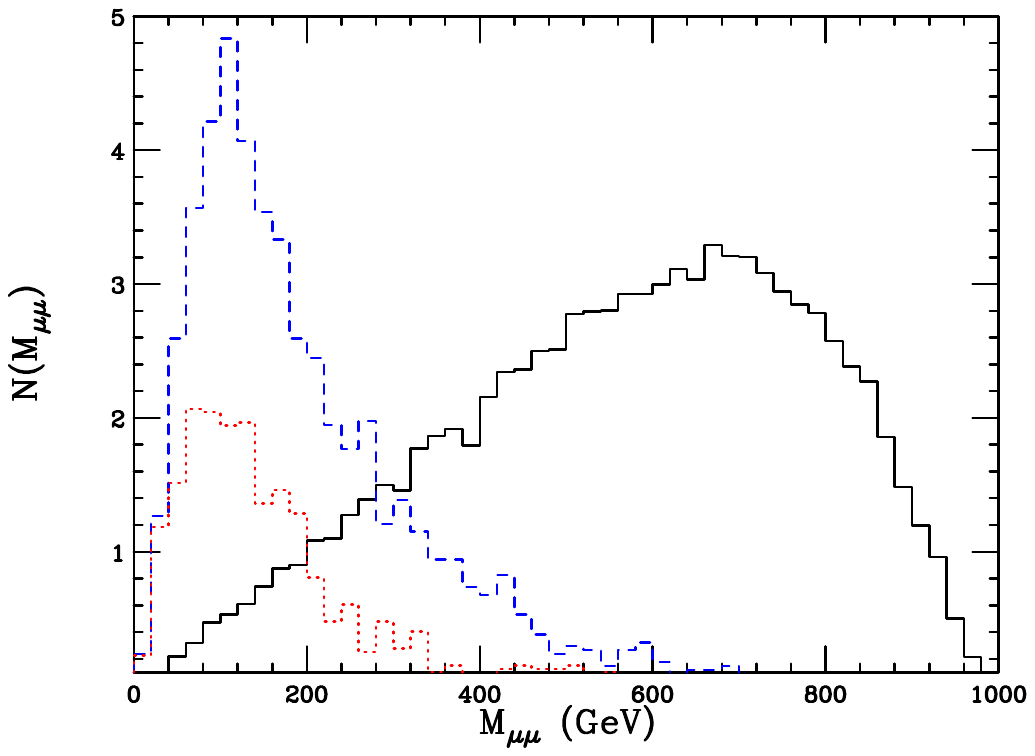}
\vspace{-1.8cm}  \caption{Distributions of the transverse momentum of the hardest muons
    (left) and of the same-sign muon invariant mass (right).
    The solid histograms are the signals, the dashes correspond to 
    four tops, the dots to the $bbZZ$ background. }
  \label{fig3}
  \end{center}
  \end{figure}

Several observables were presented in \cite{cccf3} for the purpose of the
benchmark BM I: as for leptonic decays, in Fig.~\ref{fig3} the hardest
muon transverse momentum ($p_{T,1}$) and the invariant mass of same-sign
muons ($M_{\mu\mu}$) are plotted, for both signal and background.

The background $p_{T,1}$ spectra are substantial only at low transverse momenta,
peak about $p_{T,1}\simeq 100$~GeV and
rapidly vanish at large $p_T$, while the
signal ones are broad and substantial
up to $p_{T,1}\simeq 2$~TeV.
Above a few hundred GeV, the signal greatly dominates over the background:
this was expected, since signal muons are 
related to the decay of a TeV-scale resonance.
Regarding 
$M_{\mu\mu}$, unlike the
backgrounds, whose spectra peak at
low values and are negligible above 500 GeV,
the signal yields a broad invariant-mass spectrum,
shifted towards large values and exhibiting a maximum about 700 GeV.
\vspace{-2.5cm}

\section{Discussion}
I reviewed recent phenomenological work carried out
in Refs.~\cite{cccf1,cccf2,cccf3} within the
$SU(3)_C\times SU(3)_L\times U(1)_X$, a.k.a. 331 model,
which has, above all, the nice features to explain the number of quark and
lepton families and the asymmetry between the third and the first
two quark families. I worked in the framework of \cite{PHF}
and explored the possibility to discover doubly-charged
bileptons, i.e. doubly-charged vectors or scalars with lepton number $\pm 2$,
at the LHC and at a future 100 TeV collider FCC-$hh$.
I first considered production of bilepton pairs and decays into same-sign
leptons, then I accounted for non-leptonic decays too.
In both cases, a few benchmarks, not yet excluded by the experimental searches,
were determined in such a way to maximize the cross section at LHC.
In the case on leptonic decays, it was found that a discovery of
bileptons is feasible, with a possible signal due to vector bileptons
dominating over the scalars, because of helicity arguments.
Decays into non-leptonic final states are more cumbersome,
 since bileptons decay into a heavy TeV-scale quark and a light Standard Model
quark: the cross section of the resulting decay chain is too small
at the LHC, even in the high-luminosity phase, to give any signal.
Nevertheless, non-leptonic decays of bileptons are expected to be visible at
a future FCC-$hh$.

Before concluding, I wish to stress that,
while the work here presented deals with primary
production of bileptons, it is certainly very interesting
a scenario with the TeV-scale quarks $T$, $S$ and $D$ heavier than
$Y^{++}$, so that they can be produced in processes like $pp\to T\bar T$
and decay according to, e.g., $T\to Y^{++} b$. A study of 
heavy-quark production and decays, along the lines of \cite{panizzi},
but specific to the 331 model, is currently
in progress \cite{prog}.
 In summary, as the most studied models of new physics have given no
visible signal yet, exploring alternative scenarios is compelling.
The bilepton model is certainly an appealing framework from the
theoretical viewpoint and, as summarized in this contribution, it
features a rich phenomenology which may be a first indication of new physics
in the next LHC run as well as HL-LHC and FCC-$hh$. It is then
hopeful and desirable that the experimental collaborations use the results here
presented and join the effort to search for bileptons at present
and future colliders.

\vspace{6pt} 

\funding{This research received no external funding.}
  
\dataavailability{The results presented in this paper are openly available
  in Refs.~\cite{cccf1,cccf2,cccf3}. The release of the computing codes used to obtain
  such results is under way. For the time being, the codes are available on request
from the author.} 

\acknowledgments{It has been a great honour collaborating with
  Paul Frampton, who introduced me to the 331 model and gave me countless
  hints on physics beyond the Standard Model. I also acknowledge Antonio
  Costantini and Claudio Corian\`o, coauthors
  Refs.~\cite{cccf1,cccf2,cccf3}.}

\conflictsofinterest{The author declares no conflicts of interest.}



\reftitle{References}

\PublishersNote{}
\end{document}